# A magnetic excitation linking quasi-1D Chevrel-type selenide and arsenide superconductors

Logan M. Whitt[1], Tyra C. Douglas[1], Songxue Chi[2], Keith M. Taddei[2], Jared M. Allred[1]

[1] Department of Chemistry and Biochemistry, The University of Alabama, Tuscaloosa, 35487-0336

[2] Neutron Scattering Division, Oak Ridge National Laboratory, Oak Ridge, Tennessee 37831, USA

**Abstract**

The quasi-one-dimensional Chevrel phases, $A_2Mo_6Se_6$ ($A$ = Tl, In, K, Rb, Cs), are of interest due to their atypical electronic properties. The Tl and In analogues undergo a superconducting transition whereas the alkali metal analogues show charge gapping of another, not well understood type. We report the results of inelastic neutron scattering on polycrystalline $In_2Mo_6Se_6$ ($T_c$ = 2.85 K) and $Rb_2Mo_6Se_6$ (non-superconducting) samples, which reveal a column of intensity with linear dispersion from [0 0 1/2] to [0 0 1] in both compounds. The observed temperature and $|Q|$ independence together suggest the presence of unconventional carriers with a spin contribution to the excitation. This is contrary to the prevailing model for these materials, which is that they are non-magnetic. The excitation has similar dispersion and $S(Q,E,T)$ behavior as one observed in the structurally related superconducting compounds $A_2Cr_3As_3$ and $A_2Mo_3As_3$ ($A$ = K, Rb, Cs), which has been interpreted as magnetic in origin and related to Fermi surface nesting. The connection is unexpected because the calculated Fermi surface of the arsenides differs substantially from the $A_2Mo_6Se_6$ compounds, and many consider them distinct classes of materials. The new observation suggests a hidden link in the physics between both classes of superconductors, perhaps originating from their quasi-low-dimensional character.

**Introduction**

Recently, two important families of superconductors, the condensed Chevrel phases and chromium pnictides, became linked with the discovery of $A_nCr_3As_3$ ($A$ = K, Rb, Cs; $n$=1,2,)[1-6] and later $A_2Mo_3As_3$ ($A$ = K, Rb, Cs)[4, 7, 8]. The condensed Chevrel phases are a subgroup of the more common Chevrel phase family of superconductors and have the form $A_2Mo_6X_6$ ($A$ = In, Tl, alkali metal, $X$ = S, Se, Te).[9-13] Structurally, the latter family consists of directly-bonded $Mo_3$ triangles capped by edge-bridging $X$ atoms. This approximately planar $[Mo_3X_3]^{1-}$ motif stacks to form infinite chains, which are separated from one another by $A$-site cations (Figure 1a). The crystal structure is represented in the $P6_3/m$ space group type, and the lack of direct bonding interactions between chains is why the compounds are considered quasi-one-dimensional (quasi-1D). Properties are similar between different $X$ choices, but the sulfide and telluride analogs are not as thoroughly characterized as the selenide analogs. This article will focus on the latter class.

The arsenide analogs contain the same structural motifs and the $n$ = 1 versions are isostructural to the selenides. Note that the original Chevrel compounds historically followed a different nomenclature, and so their stoichiometries should be divided by two when comparing to the arsenide compounds. The $n$ = 2 compounds have twice as many $A$-sites between chains, giving a slightly different crystallographically averaged symmetry, shown in Figure 1b and represented by the space group type $P\bar{6}m2$. The $n$ = 2 compounds are the superconducting arsenides, and they contain about 0.67 fewer electrons per metal atom than the traditional condensed Chevrel phases,[11, 12, 14] leading to a large shift in the expected Fermi surface and presumably physical properties.[15] For comparison, an additional 0.67 electrons per metal atom beyond $A_2Mo_6Se_6$ shifts the properties to strongly ferromagnetic in the form of $TlFe_3Te_3$ ($Tl_2Fe_6Te_6$).[16] Given that it has even been proposed that the isoelectronic Mo and Cr arsenides have different superconducting mechanisms,[8, 17] it is a reasonable assumption that the selenides also differ substantially from the arsenides. Nevertheless, superconductivity is prevalent across both Se and As subtypes, and there are still many questions about which physical properties can be considered analogous, and which are merely coincidental.

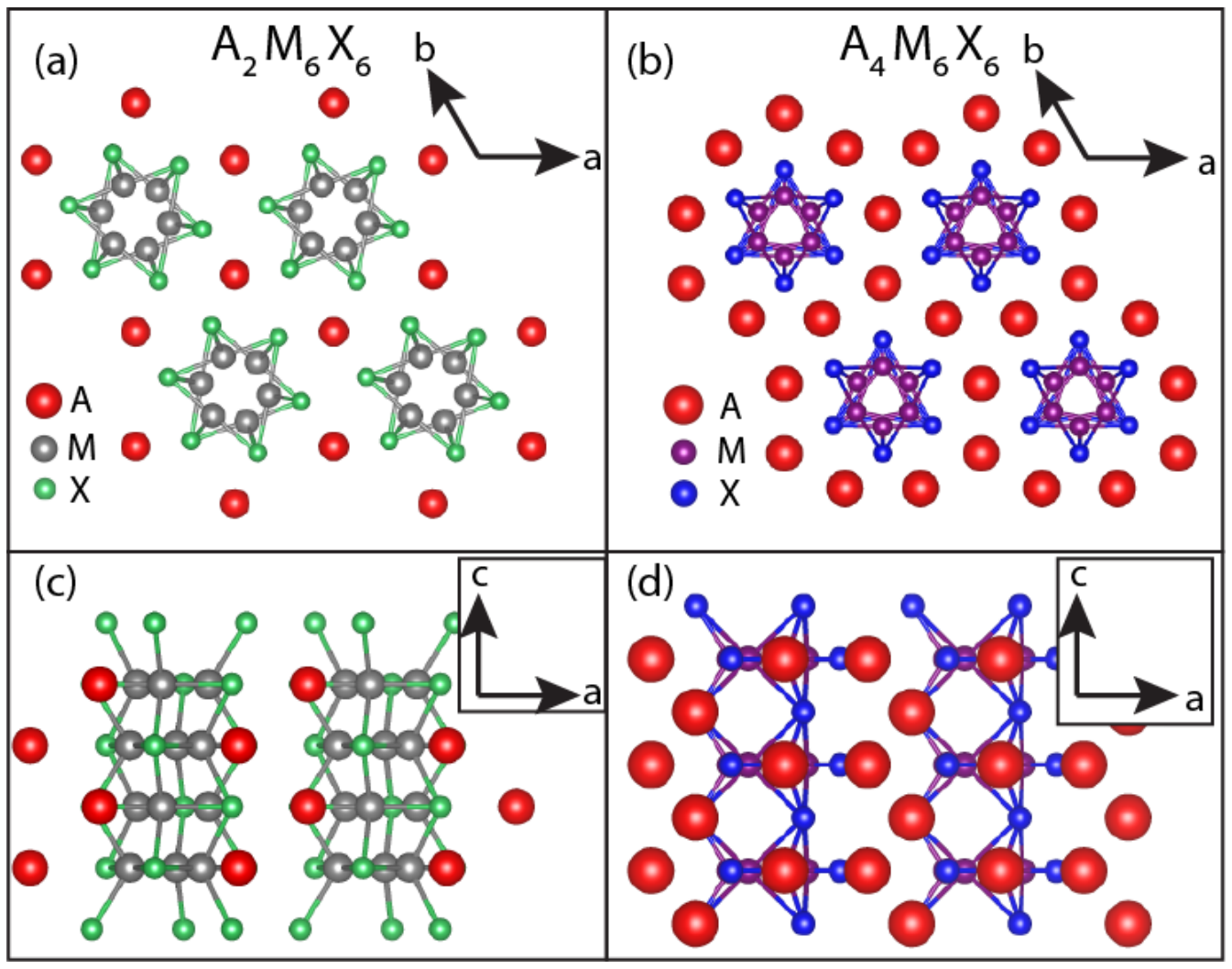

Figure 1: (a) Crystal structure for $A_2Mo_6X_6$ ($A$ = Tl, In, Rb, Cs; $X$=S, Se, Te) and $ACr_3As_3$ ($A$ = K, Rb, Cs,) viewed along the c-axis. (b) Crystal structure for $A_2M_3As_3$ ($A$ = K, Rb, Cs; $M$=Cr, Mo) viewed along the c-axis. (c) Same structure as (a) viewed along the b-axis. (d) Same structure as (b) viewed along the b-axis.

Perhaps the most notable phenomenon separating the arsenides, especially $A_2Cr_3As_3$, from the selenides is magnetism. Nuclear magnetic resonance and muon spectroscopy studies on the arsenides revealed significant electron correlations as well as magnetic fluctuations even in the absence of any observed long range magnetic ordering.[18-20] A magnetic instability resembles the physics of the iron-based superconductor (IBS) family as well as other families of unconventional superconductors.[21] It is this interpretation that links the physics of the chromium arsenide type condensed Chevrel phases to the structurally distinct IBS class. This is in contrast to the traditional condensed Chevrel phases, which are known to contain an electronic instability that has usually been attributed to a charge density wave (CDW) ($A$ = K, Rb, Cs) with a subtle metal to insulator transition (MIT) component, which can be suppressed in favor of superconductivity ($A$ = Tl, In).[22-24] It is worth noting that the $A$-site ion is believed to play a crucial role in the resultant properties of these compounds.

DFT and LDA calculations performed by Petrović *et al.* show that the In,Tl contribute substantially to the bands near the Fermi level whereas the alkali metal analogues have almost no contribution to these bands.

Further analogies between chromium pnictides IBSs have been encouraged by inelastic neutron scattering measurements on polycrystalline $K_2Cr_3As_3$ which uncovered a column of excitations originating around $Q$ = 0.7 $Å^{-1}$, which is indexable to $\left(00\frac{1}{2}\right)$ .[25] The excitation can be accurately modeled as magnetic modes originating from $\left(00\frac{1}{2}\right)$ and $\left(10\frac{1}{2}\right)$. This is interpreted as "incipient magnetism" because there is no evidence of magnetic ordering in the bulk properties or via neutron diffraction. Static, short-ranged symmetry-breaking correlations have also been observed in the chromium arsenide phases by PDF and DFT analysis, albeit with a distortion direction that is distinct from the observed excitation attributed to magnetism suggesting the chromium pnictides are near both magnetic and structural instabilities.[26] However, while such instabilities have been both experimentally observed and theoretically predicted via electronic structure calculations in the chromium pnictides, no equivalent phenomena have been observed in the Chevrel phases and the results of electronic structure calculations show no equivalent structural instability.[22, 27] Moreover, no similar measurements sensitive to short-ranged magnetic or structural correlations have been performed on the selenides, to our knowledge, complicating attempts to draw a comparison.

Like the chromium arsenides, the selenide phases are reported to show no evidence of magnetism in the bulk magnetization.[14, 22] Magnetic instabilities had not been given as much attention, though the potential for inherent magnetism in both superconducting and non-superconducting selenide Chevrel phases has been commented on for at least 30 years.[24]. More recently, a detailed study of their physics found evidence for strong electron correlations and determined that all $A_2Mo_6Se_6$ compounds, including the superconducting ones, were on the verge of an SDW transition. Even so the SDW hypothesis was not given further consideration due to the lack of physical properties supporting magnetic ordering. Instead, a CDW or Peierls-like scenario that included rattling guest ions was given the most weight. Notably, subsequent RIXS experiments weakened this model somewhat by finding little evidence in support of a Peierls soft-phonon instability and by invalidating the rattling guest ion hypothesis.[27]

In this light, incipient magnetism in the molybdenum selenides seems much more probable than it once was. Despite the detailed characterization of the selenide-based condensed Chevrel phases, the assumption that they are non-magnetic has precluded investigations that might prove otherwise. For example, INS has been performed on the same selenide phases, but the $E(Q)$ range was optimized for phonon measurements, and the magnetic excitation in the chromium arsenide would not have been observed.[28, 29] Additionally, several phonon modes were recently mapped out in detail via single crystal resonant inelastic x-ray scattering (RIXS), but this is not sensitive to magnetic excitations.[27]. To fill this gap, we report an INS measurement on $In_2Mo_6Se_6$ and $Rb_2Mo_6Se_6$ across the same $E(Q)$ range that was performed on the chromium arsenide phases. The experiment uncovers a column of excitations with nearly identical *E*(*Q*) dispersion in both superconducting and non-superconducting selenide analogs. The temperature, energy, and Q dependence of the newly measured excitations indicate the presence of unconventional spin carriers. This is the first direct evidence of magnetic interactions in the selenide subtype, and it links their properties to some of the same unconventional physics observed in the arsenide subtypes.

**Methods**

Polycrystalline samples of $In_2Mo_6Se_6$ and $Rb_2Mo_6Se_6$ were synthesized using the methods reported by Tarascon *et al.* [30] Specifics on the synthetic process can be found in the supporting information. Inelastic neutron scattering experiments were carried out on the HB-3 Triple-Axis Spectrometer at Oak Ridge National Laboratory's High Flux Isotope Reactor. In order to maximize the flux, horizontal collimation settings of 48-60-60-120 were used in this experiment. PG(002) was used at the monochromator and analyzer due to its high neutron reflectivity. The analyzer energy was fixed to 14.7 meV for all experiments and data was collected by performing constant energy transfer scans over low *Q* regions. The $In_2Mo_6Se_6$ compound was analyzed at 3,5,7,9,and 12 meV at 5 and 300 K. $Rb_2Mo_6Se_6$ was measured at the same energies at 5 K. Resolution function analysis was performed using the DAVE software package.[31]

**Results**

Figure 2(a) shows a neutron intensity map of the dynamic structure function $S(Q,E)$ for $In_2Mo_6Se_6$ at 5 K. This is above the superconducting transition in this compound, which allows the comparison between compounds in their normal states. Temperatures below $T$c were not accessible during this experiment. A very broad column of excitations is clearly observed between $0.80 \leq Q \leq 1.45$ Å$^{-1}$. This feature is also seen in the 300 K scans for the same compound (Fig. 2(b)). The column does not appear to have any gaps within the energy resolution of the measurement.

The 300 K and 5 K scattering intensities are quite similar, which is shown in the representative *Q*-cut at 9 meV shown in Fig. 2c. Despite the excess intensity at 3 meV in the 300 K measurements, the main peak itself shows no statistically significant change in intensity between measurements at any energy. This is surprising, since common excitations including magnons and phonons typically show enhanced intensity scaled by $\left(1 - e^{-E/k_B T}\right)^{-1}$ above kT, which is below 150 K for all energies measured here. The only portion of the measured scattering that scales with the Bose factor is on the high-Q end of the cuts, and can be assigned to phonons based on prior work[27]. More details showing how the phonon assignment is made can be found in the supporting information. The 5 K INS data for the rubidium analogue is presented in Fig. 3. The spectrum shows a very similar $S(Q,E)$ as the $In_2Mo_6Se_6$ at the same temperature. The subtle differences between the two is addressed at the end of this section.

The $\mathbf{Q}(E)$ dependence of the INS peak in $In_2Mo_6Se_6$ was determined in several steps. First, the general trend of $|\mathbf{Q}(E)|$ was estimated by fitting the scattering cuts to a gaussian curve with constant background per scan. Since the samples are polycrystalline, the experimental conditions also require a Lorentz factor correction of $1/\sin(2\theta)$ to be applied to the peak intensity. Since the orientation of $\mathbf{Q}$ is not known for these initial fits, only the projected dispersion can be estimated, which is about 0.06 Å$^{-1}$ per meV and approximately linear. This approximate dispersion was used to determine the effect of the instrumental resolution on the measurement.

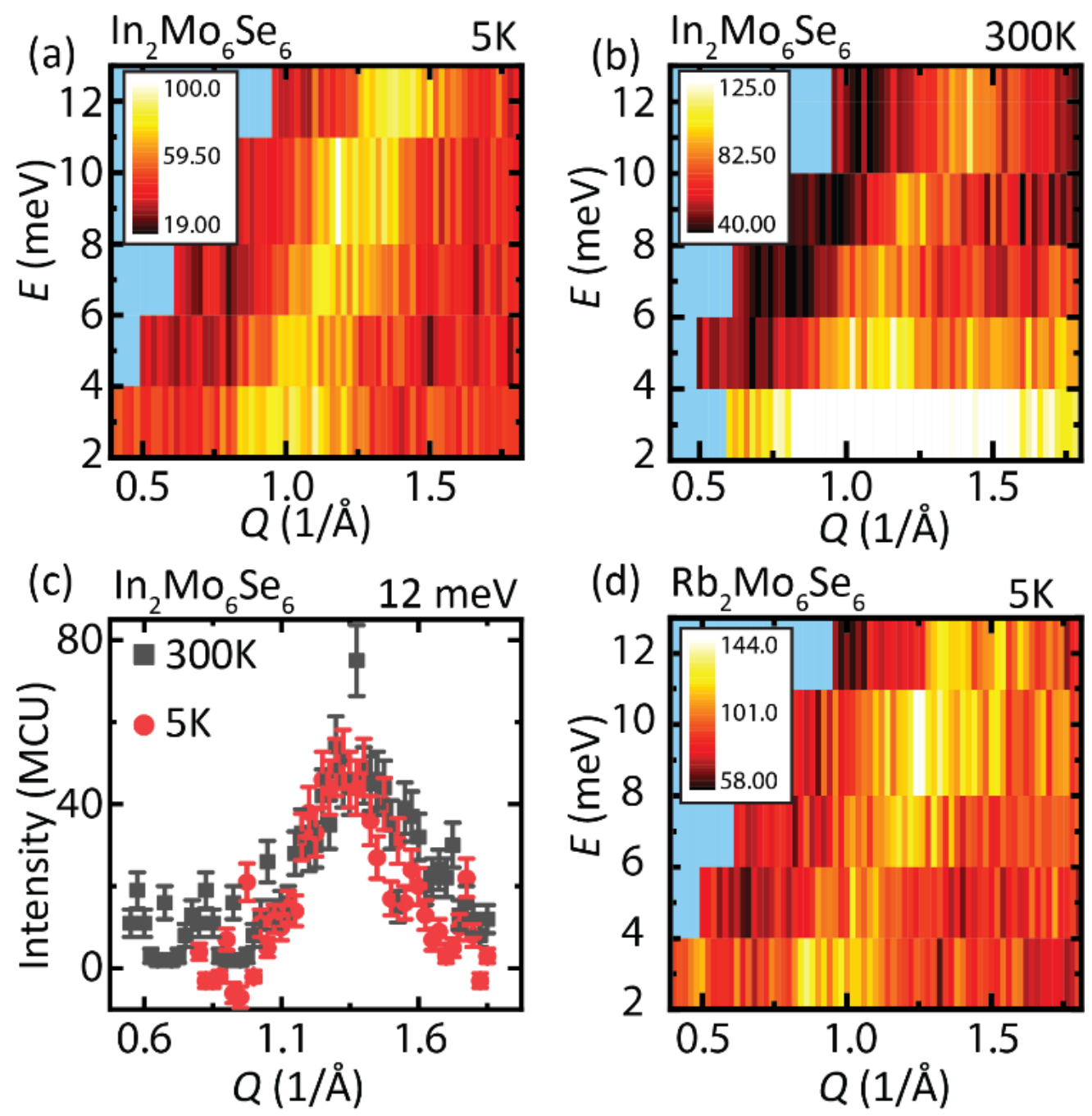

Figure 2: Inelastic neutron spectra intensity maps for $In_2Mo_6Se_6$ at (a) 5K and (b) 300K. (c) Comparison between 5 and 300 K plots at 12 meV. (d) Inelastic neutron spectra intensity maps for $Rb_2Mo_6Se_6$ at 5K. Intensities in all panels are relative to the monitor count units divided by 300 (MCU). No additional scaling or background subtraction is applied.

A higher order harmonic peak was identified in the $In_2Mo_6Se_6$ sample at 5 K by making a separate $Q$-cut up to $Q$ = 2.4 Å$^{-1}$ at 3, 5, and 7 meV (Fig. 3a-c). The raw intensities of the extended cut (black circles) show a broad hump at higher $Q$ that has a negative $Q/E$ dependence, consistent with a higher order harmonic. The overlay (orange line) shows that the intensity is well described in all three energies using two identical peaks. Only one new parameter is introduced to the model from the initial model: the position of the higher order peak in $Q$. The deconvoluted form of the two-peak model is shown as the offset blue line in the same figure panels for each energy. The observed difference in peak shape arises from the ellipsoidal shape of the instrumental resolution function (see supporting materials for more details). The result of these fits are presented in Fig. 3d. The midpoint of the first and second order peaks is the same within error for each energy (1.53(3) Å$^{-1}$, green triangles in Fig 3d), which confirms that the peaks are -/+$Q$ branches of the same mode. The fitted peak gaussian widths, σ, are all equivalent within error, with an average of 0.365(19) Å$^{-1}$.

The purpose of the modeling summarized in Fig. 3d is to determine the relationship between the observed signal and the compounds' actual structures. Of utmost importance in this determination is the critical points in the Brillouin zone that are traversed, which requires the orientation of $\mathbf{Q}$, which is difficult to index in polycrystalline samples. In the present case it is possible to index the excitation, because most reciprocal space axis choices yield inconsistent dispersions when applied to the first and second order peaks. In fact, dispersion along the [0 0 $Q_L$] axis is the only option that gives consistent dispersions for both first and second order peaks. The revised 3, 5, and 7 meV dispersions shown in Fig 3d are well fit by a linear model, with $R^2$ = 99.5% and 99.8% for the first and second order peaks, respectively. The linear fits intersect zero energy transfer at $Q_L$ = 0.615(9) and 1.580(6) for the first and second order peaks, respectively, and converge at $Q_L$ = 1.10(2) and $E$ = 13.0(5) meV. The suggests an apparent first and second order excitation dispersion of [0 0 ½] to [0 0 1] and [0 0 3/2] to [0 0 1], respectively, with a systematic offset of about +0.1**c***(0.14 Å$^{-1}$) between expected and apparent peak center across all observations. This is likely a systematic error that originates from the use of a simplistic gaussian peak shape that lacks peak asymmetries specific to excitation models.

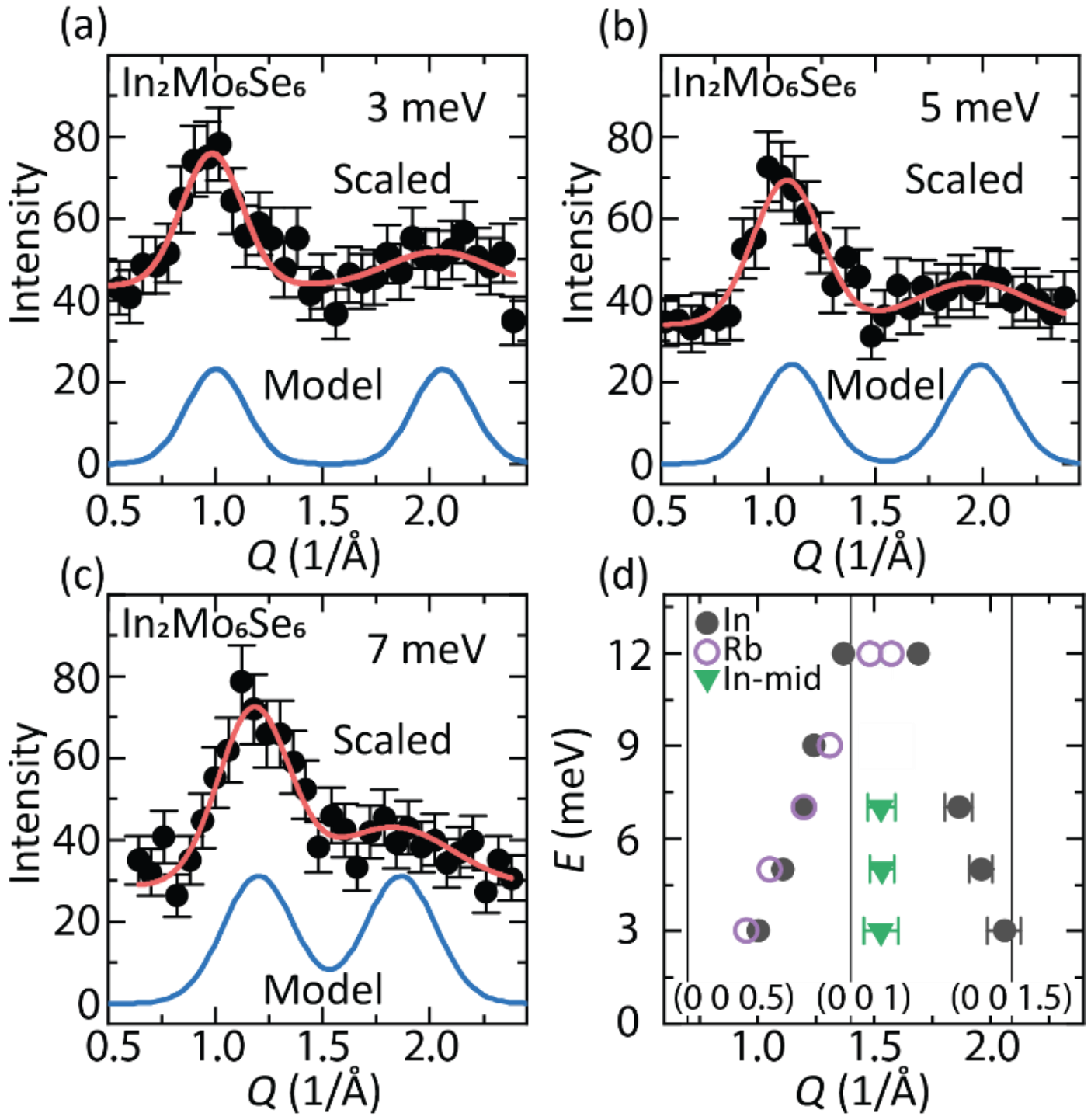

Figure 3: (a-c) Extended $Q$ cuts of $In_2Mo_6Se_6$ at 5 K and 3, 5, and 7 meV, respectively. The blue curve shows the fitted two peak model and the red overlay is the same model scaled according to experimental conditions. (d) $E(Q)$ dispersion for $In_2Mo6Se_6$ (closed symbols) and $Rb_2Mo_6Se_6$ (open circles). Green triangles represent the midpoint of two independently refined peak positions. The vertical lines represent indices of special points.

Based on this linear dispersion, it is likely that the 1st and 2nd order peaks are overlapping at 12 meV. In order to estimate the peak locations at this energy, a constraint was added to the peak shape model that forces the two peaks to center around $Q$ = 1.53 Å$^{-1}$. When this constraint is applied to the 12 meV model, the observed gaussian width, σ, matches the determined width for the 3 to 7 meV fits (0.36 Å$^{-1}$) within error. While this suggests that the model is at least qualitatively applicable at this energy, it is likely that it overestimates $Q_2$ – $Q_1$ due to the convolution of the fronted peak shapes. To minimize such parametrical correlations, the σ term was constrained to the average from the other energies in the final 12 meV model. As for the 9 meV model, the majority of the second order peak was outside the measured range. The constrained midpoint model can be used to better model the visible tail of the second order peak, though doing so has an insignificant effect on the first-order portion of the fitted model.

$Rb_2Mo_6Se_6$ dispersion is also shown in Fig. 3d (open circles). Only the edge of the second order peak is visible in all of the $Q$ cuts except for at 12 meV, where the two peaks overlap. The two peak model with a constrained midpoint was used in all of the Rb models to account for the upturned intensity at higher $Q$ that arises from the leading edge of the second order harmonic. The 12 meV cut was modeled in an identical fashion in both compounds. The same 1.53 $Å^{-1}$ midpoint was used for all Rb cuts because the $c$ cell axis of the Rb and In analogs are equivalent within 0.3%. This analysis shows that the Rb analog's excitation is slightly less dispersive than the In one, and it converges at the zone boundary at a slightly lower energy. Together this suggests a slightly weaker interaction in $Rb_2Mo_6Se_6$ than in $In_2Mo_6Se_6$. Though the Rb analog measured molar intensities range from 0.7 and 1.1 times the In ones, the difficulties in modeling the background's energy dependence makes this quantity most susceptible to systematic error in the model. Most conservatively, it can be stated that the measured intensities are the same within a factor of two according to the present measurement.

**Discussion**

Direct comparison of the observed molybdenum selenides phases' excitations to those of $K_2Cr_3As_3$ in Ref. [25] reveals some clear similarities. The column of excitation is in nearly the same region of $Q$-space and with similar $Q$ dispersion for all of these compounds. The main differences are that the selenides in the present study appear to have slightly sharper and more intense features. Despite the close similarities, it is possible that the exact origin of the excitations differs between compounds. To explore this topic, the discussion considers various phenomena observed or proposed in the selenide and arsenide families.

Magnetic ordering is visible to neutrons through a structure factor that depends both on the moment component perpendicular to $\mathbf{Q}$ and the length of $|\mathbf{Q}|$. Since the measured excitation is interpreted to fall along $(0, 0, Q_L)$ direction, the orientation contribution is expected to be constant. The $|\mathbf{Q}|$-dependent portion of the form factor decreases with Q, but the change is too weak to have a significant effect on the model shown in Fig 3a-c. This means that the observed signal in $In_2Mo_6Se_6$ is entirely consistent with the expected magnetic structure factor for Mo. Additionally, the observed fronted peak profile is consistent with magnetic peak profiles such as those used to model the related $K_2Cr_3As_3$ excitation,[25] which can explain the systematic error in peak location of about $+0.1\mathbf{c}^*$ in the simple gaussian model used here. These observations are all consistent with a magnetic form factor contributing to the scattering. However, magnon intensities should scale with temperature via the Bose factor, yielding *ca.* 9 to 2 times increased intensity at 3 and 12 meV, respectively. This is not observed (Fig. 2a-c). Also, magnons that originate from $\mathbf{Q}_0 = (0, 0, L/2)$ would be expected to also have measurable magnetic ordering peaks consistent with a $\mathbf{k} = \left(0 0 \frac{1}{2}\right)$ origin, yet none are observed in the elastic scattering pattern on HB-3 for either of the compounds at 5 K. This suggests a more complex origin of the excitation.

For On the other hand, a conventional phonon model can be effectively ruled out from consideration. Phonons have already been investigated in $A_2Mo_6Se_6$ for $A$ = K, Rb, Cs, In, and Tl.[22, 27, 29] Most recently,

Gannon *et al.* performed a detailed characterization of phonon modes along the $\Gamma$ to $A$ ([$H$ $K$ $L$] to [$H$ $K$ $L/2$]) reciprocal space points using single crystal RIXS, along with calculations consistent with the observations.[27] None of the measured or calculated phonons show a dispersion consistent with the observed excitation. Additionally, the characteristic $|\mathbf{Q}|^2$ scaling of a phonon's $S(\mathbf{Q})$ is missing. These points, combined with the lack of Bose-like temperature statistics, all strongly contraindicate a conventional phonon excitation. For comparison, the related excitation in $K_2Cr_3As_3$ was found to be inconsistent with a phonon for similar reasons.[25]

Ruling out conventional phonons and identifying a spin component to the scattering is sufficient evidence to conclude that the observed excitation is incompatible with the prevailing model governing the molybdenum selenide condensed Chevrel. The remaining discussion focuses on examples of less conventional phenomena drawn from the current literature on the arsenide family, and considers how consistent they are with the present observation.

The enhanced intensity of conventional phonons and magnons above $k_BT$ is associated with the thermal population statistics for non-conserved boson quasiparticles.[32] These constraints tend to break down under special conditions such as in low-dimensional electronic systems. Tomonaga-Luttinger Liquid (TLL) physics, wherein spin and charge are carried by separate, conserved quasiparticles, is a well known example of unconventional electronic behavior in a 1D system [33]. In fact, TLL physics has previously been applied to $K_2Cr_3As_3$ [20, 34, 35] and $Rb_2Mo_3As_3$[8]; the latter was interpreted as a very rare example of a TLL with attractive interactions between quasiparticles. The apparently gapless, linearly dispersive INS spectra reported here for selenide condensed Chevrel phases are reminiscent of spin dynamic spectra in known TLL systems such as $BaCo_2V_2O_8$[37], and the temperature independent INS intensity can be attributed to the different quasiparticle population statistics.

From this initial analysis, the present INS spectra are consistent with a TLL scenario, and even have features supportive of that interpretation. The connection should nonetheless be treated as speculative, since it is the power-law decay of intercarrier correlations that is considered the characteristic feature of TLL physics. The charge transport properties of the molybdenum selenides only seem to show the power-law transport properties characteristic of TLL physics when isolated as nanofilaments,[36] which is a phenomenon that is either not present or obscured in the bulk [22]. This is not sufficient to discard the model, though, since the same is true of the molybdenum and chromium arsenides. These compounds instead show power-law dependence in, for example, the nuclear spin relaxation rates, which is probed by NMR. [8, 38]

In addition to TLL physics, other recent examples of unconventional electronic properties or hidden ordering in related compounds include geometric frustration of displacements in $K_2Cr_3As_3$,[26] purported topological superconductivity ($Tl_2Mo_6Se_6$ and $(Na,K,Rb,Cs)_2Cr_3As_3$),[39-42] and missed hydrogen intercalation in nominally $K_2Cr_3As_3$ that is actually $KH_xCr_3As_3$ [43] with a potential Lifshitz transition.[44] Altogether we note a consistent trend of non-trivial electron-phonon coupling, links between spin and lattice ordering, and hidden symmetry-breaking correlations. The connection to topologically non-trivial electronic states in closely related $Tl_2Mo_6Se_6$ is worth highlighting, since similar topological considerations are linked to anomalously high spin-charge conversion in the chemically similar compound $WTe_2$ [45]. Since the listed phenomena are not all mutually exclusive, it is premature to force the present findings into a single model. Additional experiments are required to determine which, if any, of the aforementioned phenomena apply to the present compounds.

**Conclusion**

A magnetic excitation is revealed in polycrystalline $A_2Mo_6Se_6$ ($A$ = In, Rb) using inelastic neutron scattering. The excitation is present across the entire energy range measured, 3-12 meV, and converges to $\mathbf{k}$ = (0, 0, 1/2) propagation vector at zero energy transfer. The $S(\mathbf{Q})$ dependence, peak shape, and absence from inelastic x-ray scattering measurements are all taken as evidence in support of a magnetic component to the excitation. The lack of temperature dependence for the measured excitation intensity suggests unconventional physics such as TLL behavior, which has been observed in the chromium and molybdenum arsenide superconductor families.

These findings enhance the electronic property connection between the molybdenum selenide superconductors and the chromium arsenide ones. How this relates to the superconducting mechanism remains an open question, since it has been proposed that the arsenides are proximal to two separate superconducting ordering instabilities and that the chromium and molybdenum arsenides have different superconducting mechanisms.[17] Since the $A$-site contribution to the band structure of the molybdenum selenides is known to play a crucial role in their properties,[22] one might expect a difference between $A$ = In and Rb analogs. Though $In_2Mo_6Se_6$ is in its normal state at both measured temperature (5 K and 300 K), the small difference observed compared to non-superconducting $Rb_2Mo_6Se_6$ is nonetheless notable in this regard. Further characterization of the magnetic excitation reported here will be an important step toward understanding the relationship between superconductivity and other types of electron correlations in these materials.

**Acknowledgements**

The work was supported by the U.S. Department of Energy, Office of Science, Office of Basic Energy Sciences, EPSCoR and Neutron Scattering Sciences under award DE-SC0018174. This research used resources at the High Flux Isotope Reactor, a DOE Office of Science User Facility operated by the Oak Ridge National Laboratory and was supported in part by the US Department of Energy, Office of Science, Office of Basic Energy Sciences.

# Supplemental Material for:

# A magnetic excitation linking quasi-1D Chevrel-type selenide and arsenide superconductors

Logan M. Whitt[1], Tyra C. Douglas[1], Songxue Chi[2], Keith M. Taddei[2], Jared M. Allred[1]

[1] Department of Chemistry and Biochemistry, The University of Alabama, Tuscaloosa, 35487-0336

[2] Neutron Scattering Division, Oak Ridge National Laboratory, Oak Ridge, Tennessee 37831, USA

## I. Synthesis of $In_2Mo_6Se_6$ and $Rb_2Mo_6Se_6$

Polycrystalline samples of $In_2Mo_6Se_6$ were synthesized by direct reaction of the elements. Stoichiometric amounts of In (99.9% Alfa Aesar) and Mo (99.95% Alfa Aesar) powders and Se shot (99.999% BeanTown Chemicals) shot were added to an aluminum crucible and sealed in an evacuated quartz ampoule. The tube was heated at a rate of 2 ˚C /min to 600 ˚C and kept at this temperature for 24 hr. The furnace was then heated to 1000 ˚C at 5 ˚C /min and kept at this temperature for 72 hr. The samples were slowly cooled down and the black polycrystalline sample was obtained. Needle-like single crystals were also recovered using this method. Sample purity was determined using PXRD with a Bruker D2-Phaser equipped with a Lynx-eye detector and Cu-Kα radiation. Lattice parameters were determined using the Rietveld method as implemented in the GSAS-II suite.[1] The lattice parameters were determined to be $a$ = 8.834(2) Å and $c$ = 4.450(4) Å for $In_2Mo_6Se_6$ which matches closely to the reported parameters.[2] The refined occupancy for the In site is 0.900(6). This results in the formula for the In analogue to be $In_{1.8}Mo_6Se_6$, however this analogue will still be referred to as $In_2Mo_6Se_6$ for consistency.

The synthesis of polycrystalline samples of $Rb_2Mo_6Se_6$ was adopted from Ref. 12.[3] The samples were synthesized via reaction of stoichiometric amounts of previously synthesized $In_2Mo_6Se_6$ and 10% excess RbI (99.5% BeanTown Chemicals) The samples were sealed in an evacuated quartz tube and heated to 550 ˚C at 5 ˚C /min then slowly raised to 650 ˚C at 3 ˚C/hr. The samples were left at this temperature for one week. Following this time, the samples were removed and thoroughly mixed before annealing at 650 ˚C for 4 days to ensure purity. Upon cooling, black polycrystalline, $Rb_2Mo_6Se_6$ ($a$ = 9.250(3) Å and $c$ = 4.4824(9) Å) was obtained. The lattice parameters match the reported values for this compound.[4] The refined occupancy for the Rb site is. 0.967(14). Similarly, to the In analogue, the calculated Rb formula should be $Rb_{1.9}Mo_6Se_6$.[5]

## II. Peak Profile Fits

The gaussian peak profile used in the initial peak profile fits is:

$$I(Q) = \frac{1}{\sin 2\theta} \cdot \frac{A}{w\sqrt{\pi/4\ln 2}} \cdot e^{-4\ln 2\,(\frac{(Q-Q_1)}{w})^2} + I_0$$

$A$ is the gaussian amplitude, $w$ is the gaussian full width at half maximum (FWHM), $Q_1$ is the peak center position in $Q$, $2\theta$ is the angle between the primary beam and the analyzer, and $I_0$ is a $Q$-independent background term. The refinable parameters in this model are $A$, $w$, $I_0$, and $Q_1$.

The $1/\sin 2\theta$ term is a Lorentz factor correction, which depends on $Q$ and $E$ and is not refinable in the model.. In this case, the correction is used to account for the fraction of crystallites in the powder sample that reach the analyzer, which varies as a function of $2\theta$ for a fixed collimation. The same correction is not applied in a single crystal triple axis INS experiment.

The two peak model used in the extended cuts in Figure 3a-c is slightly modified to include terms accounting for the instrumental resolution function and the position of the 2nd order peak:

$$I(Q) = \frac{1}{\sin 2\theta} \cdot \frac{A}{\sqrt{\pi/4\ln 2}} \cdot \left(\frac{1.18}{w} e^{-4\ln 2\left(\frac{(Q-Q_1)}{w}\right)^2} + \frac{1}{1.67\,w} e^{-4\ln 2\left(\frac{(Q-Q_2)}{1.67w}\right)^2}\right) + I_0$$

$Q_2$ is the peak center of the 2nd order harmonic. The refinable parameters are $A$, $w$, $I_0$, $Q_1$, and $Q_2$.

The 1.18 amplitude scale on the first order peak located at $Q_1$ is an approximation that corrects for the enhanced intensity from the focusing condition. The 1.67 width scale on the 2nd order peak at $Q_2$ is an approximation that corrects for the peak broadening from the defocusing condition. The approximations were produced by numerically simulating the effect of the resolution function on the polycrystalline sample, and then fitting a gaussian model to the resulting peak shapes to estimate changes in peak width and area.

For $In_2Mo_6Se_6$, $Q_2$ and $Q_1$ were each freely refined in the 3, 5 and 7 meV extended $Q$ cuts. At 12 meV in the same compound, the two peaks overlapped and so $Q_1$ and $Q_2$ were constrained to have the same midpoint as the 3,5, and 7 meV fits (1.53 $Å^{-1}$). The $w$ term was kept fixed at 0.33 $Å^{-1}$ at 12 meV to avoid spurious correlations between it and the overlapping peak centers.

For 9 meV in $In_2Mo_6Se_6$ and all energies in $Rb_2Mo_6Se_6$ no extended Q cuts were collected, but at least the edge of the second order peak was present in each cut. Since c* lattice parameters of the Rb and In analogs are within 0.3% of each other, the peak is expected to have the same center, and the two peak version was used for all of these cuts with a midpoint constrained to 1.53. The 12 meV peaks in the Rb analog were treated the same as the 12 meV peaks in the In analog, except that the average peak width for Rb (0.27 $Å^{-1}$) was used.

### III. Magnetic Susceptibility Data for $In_2Mo_6Se_6$

The superconducting properties of $In_2Mo_6Se_6$ were determined using zero field cooled (ZFC) alternating current magnetic susceptibility (ACMS) performed on a Quantum Design (QD) Physical Properties Measurement System (PPMS). The external field was maintained at 10 Oe with a 5.0 Oe driving field at 1000 Hz(??). The ACMS was measured on heating from 1.8 K to 16 K at a rate of 0.1 K/min. The reported superconducting transition temperature for $In_2Mo_6Se_6$ is 2.85 K[5], this temperature is marked with a solid line on the figure below.

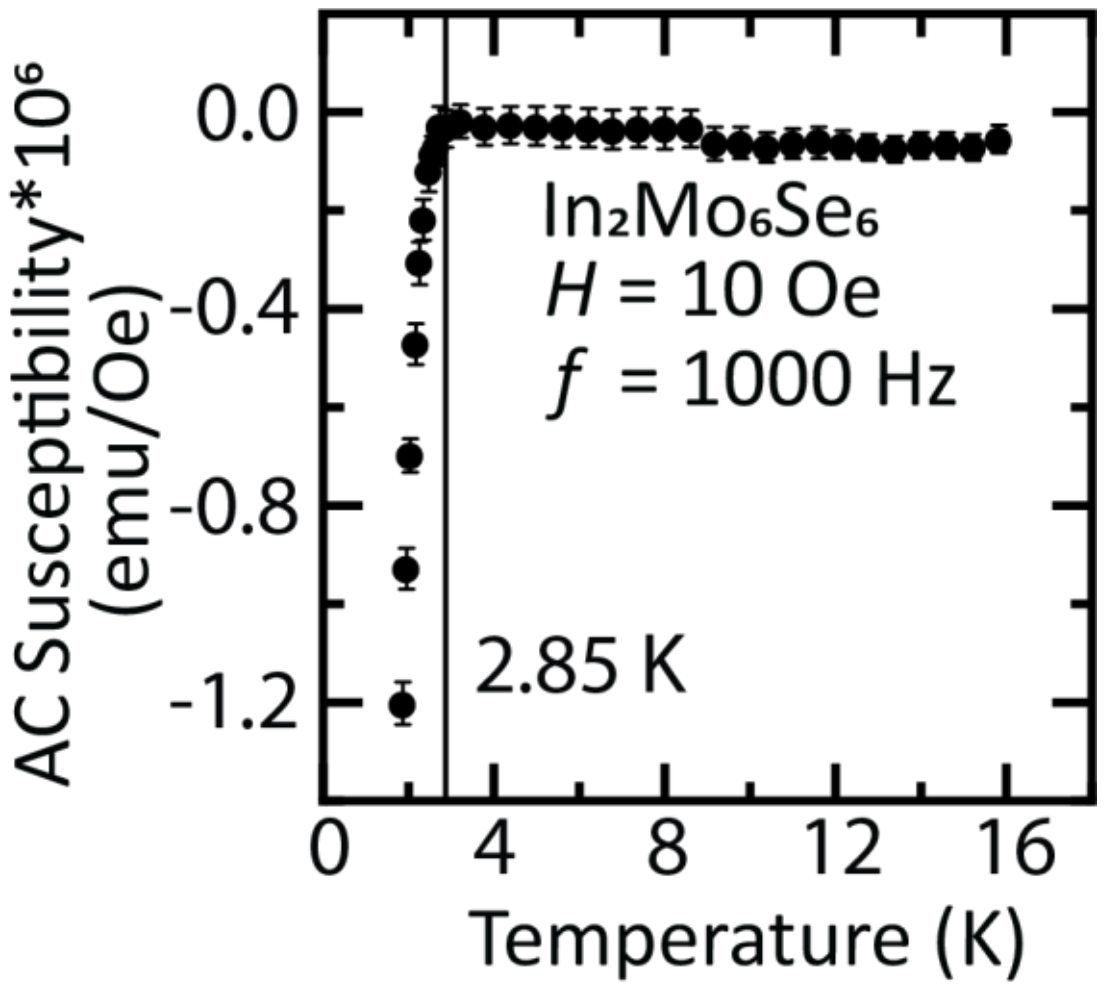

Figure S1: ZFC ACMS measurement for $In_2Mo_6Se_6$ from 1.8 K to 16 K. The previously reported superconducting transition in this compound is 2.85 K (vertical line).